\documentclass[twocolumn,epjc3]{svjour3}  

\RequirePackage{graphicx}

\RequirePackage{dcolumn}
\RequirePackage{bm}
\RequirePackage[dvipsnames]{xcolor}
\RequirePackage{subcaption}
\RequirePackage{float}
\usepackage{hyperref}
\hypersetup{
    colorlinks=true,
    linkcolor=blue,
    citecolor=blue,
    urlcolor=blue,
    filecolor=blue
}
\usepackage{doi}

\makeatletter
\def\@date{}
\def\setdate#1#2{}
\makeatother

\makeatletter
\def\makeheadbox{}%
\def\rubric{}%
\makeatother

\begin{document}

\title{CERES:\\A Cryogenic Experiment to Reconstruct\\Energy Systematics in TeO$_2$ Bolometers}

\author{Enzo Brandani\thanksref{e1,UC,lbl}
        \and
        Yael Zayats\thanksref{UC,lbl}
        \and
        Vladyslav Berest\thanksref{UC,lbl}
        \and
        Tong Zhu\thanksref{UC,lbl}
        \and
        Yury Kolomensky\thanksref{UC,lbl}
}

\thankstext{e1}{e-mail: ebrandani@berkeley.edu}

\institute{Department of Physics, University of California, Berkeley, CA 94720, USA \label{UC}
           \and
           Nuclear Science Division, Lawrence Berkeley National Laboratory, Berkeley, CA 94720, USA \label{lbl}
}

\maketitle

\begin{abstract}
Cryogenic calorimetric detectors are a powerful tool in the search for rare events such as neutrinoless double beta decay ($0\nu\beta \beta$), due to their excellent energy resolution and low intrinsic background. The performance of these detectors depends critically on a precise understanding of their energy scale and energy resolution. Recent studies suggest that both energy scale and energy resolution may vary depending on the spatial location and topology of energy deposition within the detector, indicating the presence of previously uncharacterized systematic effects.
The Cryogenic Experiment to Reconstruct Energy Systematics (CERES) is a dedicated experiment designed to directly measure the position dependence of calorimetric response in Tellurium Dioxide (TeO$_{2}$) crystals. This paper details the experimental design, current status, and future upgrade plans for CERES.

\begin{description}
\item[Keywords]
Neutrinoless double beta decay, Crystal bolometry, Low-temperature calorimeters

\end{description}
\end{abstract}

\maketitle

\section{\label{sec:Intro}Motivation}

With precise energy resolution and low intrinsic background, low-temperature crystal calorimetry (bolometry) has proven to be a robust tool for probing physics beyond the Standard Model. 
These detectors are highly reproducible and scale well for large detector arrays,
as evidenced in neutrinoless double beta decay ($0\nu\beta\beta$) searches such as CUORE \cite{CUORE:2021ctv,CUORE:2024ikf}
and its upgrade, CUPID~\cite{CUPID:2025avs}. CUORE and CUPID are instrumented with the Neutron-Transmutation-Doped Germanium (NTD)\cite{Haller1996} sensors, which are sensitive to thermal phonons. Their bolometric response has long been considered position-independent. 

The lack of  position-dependent information, however, could be a significant disadvantage of these detectors. One of the most significant sources of background for these rare nuclear event searches is degraded alpha particles from radioactive contamination of materials in the detector volume. Given the low penetration depth of these alphas ($\sim$10 $\mu$m), these events are usually confined to the surface of the detectors \cite{Kadel_2016}. 
While CUPID will eliminate the alpha backgrounds by means of particle identification through the simultaneous measurement of heat and light emission in the scintillating Li$_2$MoO$_4$ crystals~\cite{CUPID:2025avs}, surface $\beta$ events remain a significant background. 
Thus, defining a fiducial volume of the crystal that separates bulk signal events from these surface background events could significantly reduce the overall background. Moreover, topological identification of the remaining multi-site $\gamma$ events~\cite{CUPID-Mo:2023vle} could lead to the completely background-free search for $0\nu\beta\beta$ decay~\cite{CUPID:2022wpt}. 

At the level of precision of the previous experiments, 
the NTD bolometers have been assumed insensitive to event topology due to their relatively slow response time, as they are sensitive to thermal, as opposed to ballistic phonons. 
With fast sensors, such as metallic magnetic calorimeters as used in AMORE-II, a position-dependent effect on the detector response has already been observed \cite{Kim2025}. 
Additionally, a recent study of multi-crystal events in CUORE observed a small, sub-KeV effect on both the energy resolution and energy scale, suggesting there could be a measurable, albeit small, effect on the detector response due to the event topology \cite{anisha_thesis,CUORE:topology}. 
Understanding these effects in detail could improve the energy resolution of the CUORE and CUPID detectors, but also lead to the development of the position and topological reconstruction in the macro-calorimeters. 
To these ends, we present the concept and design for a novel multisensor bolometer experiment, 
\begin{figure*}[!ht]
    \centering
    \begin{subfigure}[!ht]{0.75\textwidth}
        \centering
        \includegraphics[width=.8\linewidth]{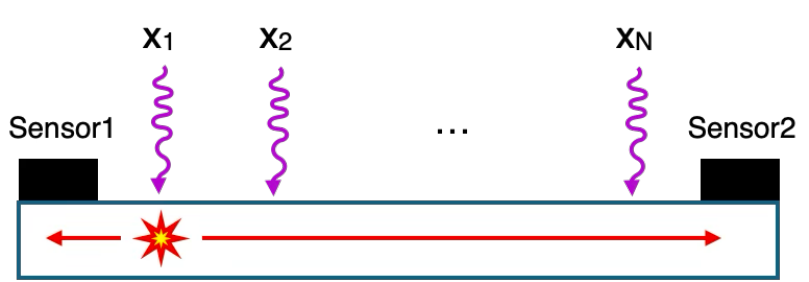}
        \caption{A 2‑D cartoon cross‑section of a 1‑D crystal depicting the CERES conceptual procedure. 
                 The blue‑bordered rectangle represents the crystal; the black squares are the thermal
                 sensors (NTD, TES, etc.); the purple squiggles are the incident radiation at a discrete
                 location $x_{N}$ along the crystal surface.}
        \label{fig:ceres_cartoon}
    \end{subfigure}

    \begin{subfigure}[b]{0.45\textwidth}
        \centering
        \includegraphics[width=\linewidth]{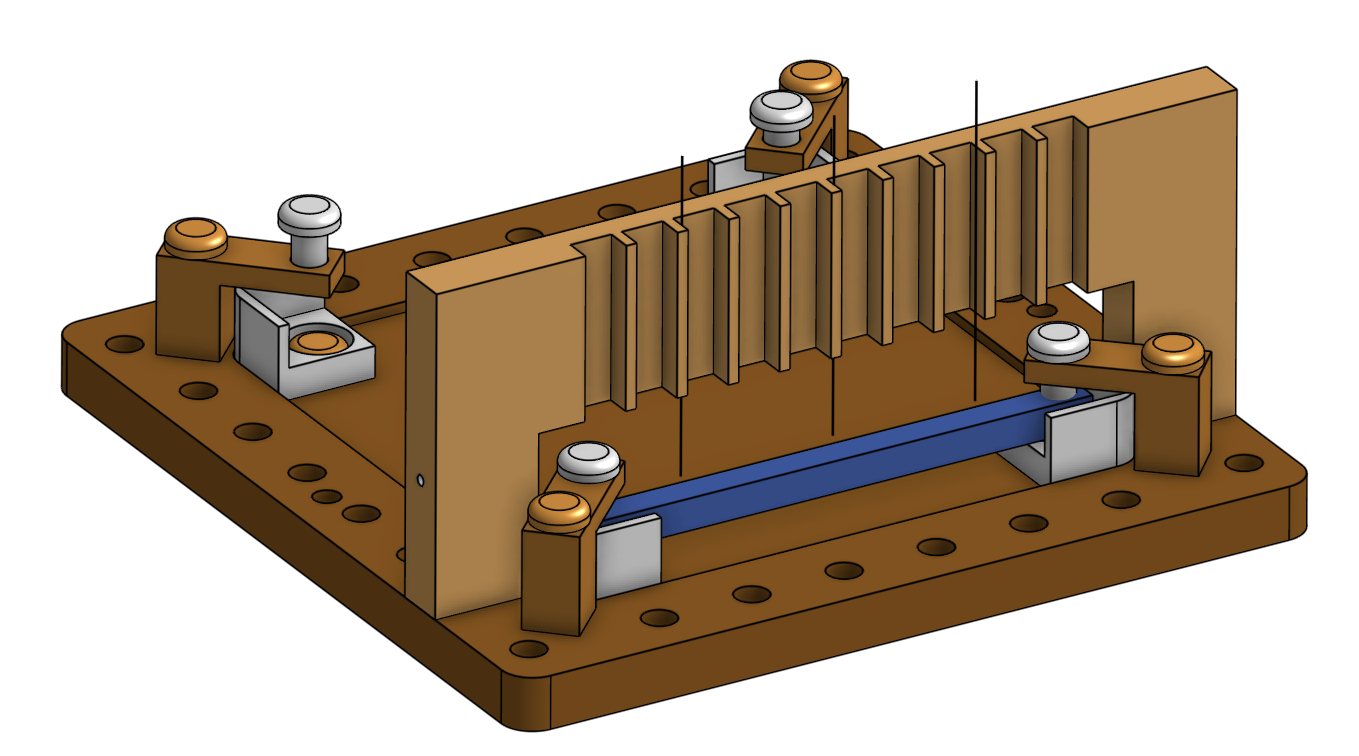}
        \caption{CAD assembly of the CERES jig, including the base, the PTFE corner pieces, the fastening towers, the harp, a dummy 1D crystal, and optical fibers arranged as they were for commissioning. }
        \label{fig:ceres_cad}
    \end{subfigure}
    \hfill
    \begin{subfigure}[b]{0.45\textwidth}
        \centering
        \includegraphics[width=\linewidth]{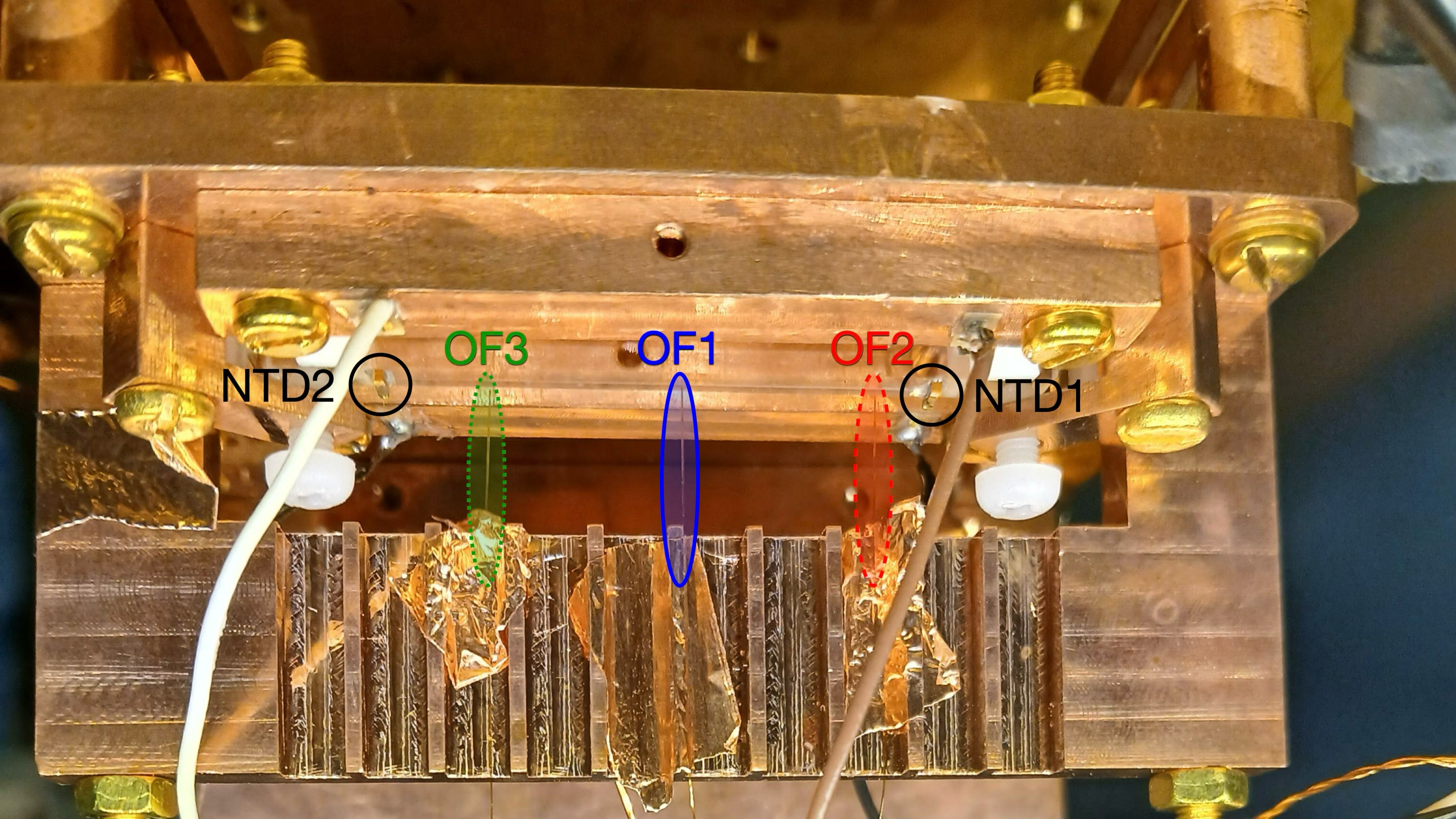}
        \caption{Photo of the CERES jig with a 1D crystal mounted to the experimental plate of the cryostat.
                 Three optical fibers, secured with copper tape to the harp, point at the crystal surface. They are labeled OF1-3 with colors and line styles corresponding to later plots. Additionally, both NTDs are circled and labeled in black.}
        \label{fig:ceres_jig}
    \end{subfigure}

    \caption{Overview of the CERES setup: (a) conceptual cartoon, (b) CAD model of the jig,
             and (c) photograph of the assembled jig.}
    \label{fig:ceres_models}
\end{figure*}

\noindent CERES, with the aim of measuring and characterizing these position and topological effects in the energy systematics of TeO$_2$ crystals.

\section{Concept and Design} 
The main goal of CERES is threefold: to measure, quantify, and parameterize the position dependence of the energy reconstruction in TeO$_{2}$ crystals.   
In designing the experiment, we target three observables: timing, amplitude, and pulse shape. 

We reduced the effective dimensionality of the crystals to perform one- and two-dimensional scans over their surfaces. Thus, we use ``strip" wafers that are 5 cm $\times$ 3 mm $\times$ 3 mm, with each pair of long faces oriented normal to [001] and [110],
and ``slab" wafers that are 5 cm $\times$ 5 cm $\times$ 3 mm, with the largest faces oriented normal to [110].

CERES uses a custom-built copper housing that allows for precise deposition of optical light at predefined positions on the surface of the crystal. 
The radiation acts as a point-like energy source, exciting phonons that stimulate an electric pulse in the bolometric sensors placed on the ends of the crystal surface. 
Currently, we opt for optical light instead of traditional silicon heaters used for calibration in similar bolometric experiments in order to produce the phonon spectrum similar to ionizing radiation and to isolate position-dependent effects inherent to the device itself, including the crystal, thermal links, and sensors. 
For devices of this scale, the heaters would contribute non-negligibly to the overall heat capacity of the system.

To enhance phonon collection, the sensors are placed closer to the site of energy deposition than the PTFE weak thermal links. 
CERES is primarily designed for the use of NTD sensors and Transition Edge Sensors (TES) \cite{TES} to convert the phonon signals to electric pulses.
One can then use the timing of the pulses to inform position reconstruction. 
Given the size of our crystals and the speed of sound in TeO$_{2}$, we expect to see timing differences in the pulses on the order of 10 $\mu$s. 
The amplitude can be used to reconstruct the relative energy sharing between sensors, which may also be position dependent considering effects due to both sensor and thermal link distance relative to energy deposition.  

In the current design, we use a 255 nm UV LED as the radiation source with an energy above the TeO$_{2}$ band gap so as to excite electron-hole pairs. 
The LED light is produced at room temperature and coupled to a single-ended optical fiber routed through the cryostat and fastened to copper pillars at each stage to ensure proper thermalization before the light is deposited onto the crystal surface. 

One challenging aspect of this experimental design is optimizing the position scans. Moving the LED source with traditional motors is infeasible at the required temperatures due to the high heat load, and while there are commercial motors rated for cryogenic use, the price encourages creative workarounds. 
In this design, the optical fibers can be secured above the crystal by insertion of a copper harp piece with multiple slots, allowing for high configurability while maintaining precise localization of the radiation.

The experimental jig used to hold the crystal wafers was an 8 cm $\times$ 8 cm square copper piece with six holes drilled along the edge of each side to mount the harp. The wafers rest on PTFE corner pieces and are secured by a ``tower" mechanism, in which copper towers with an arm hanging over the corners of the wafer hold nylon screws that clamp the wafer to the PTFE corners. The full set of pieces designed for the jig can be seen in Figure \ref{fig:ceres_cad}, and an example of the design in use can be seen in Figure \ref{fig:ceres_jig}. 
\vspace{-10pt}
 
\section{Initial Commissioning}
The first validation run of CERES was performed with two NTDs mounted with Araldite epoxy adhesive to the face normal to [001] of a strip crystal. This coupling is similar to the one employed in CUORE. Three optical fibers (OF) were mounted to the experimental jig in a symmetric configuration, with the first (OF 1) pointed at the center of the crystal and OFs 2 and 3 closer to NTDs 1 and 2, respectively. 
Measurements were taken at 15 and 16 mK in an Oxford Instruments Triton 400 dilution refrigerator with a cryogen-free pulse-tube cooler with a base temperature of $\sim$10 mK. 
To prevent destabilizing the NTD baseline voltage, data were taken with the pulse tube coolers off for a period of 250 seconds, and the LED pulsed at 4 Hz to collect sufficient statistics.
The NTDs were biased to the same resistive working point at around 1 M$\Omega$.

\begin{figure*}[!t]
\centering
  \includegraphics[width=.75\textwidth]{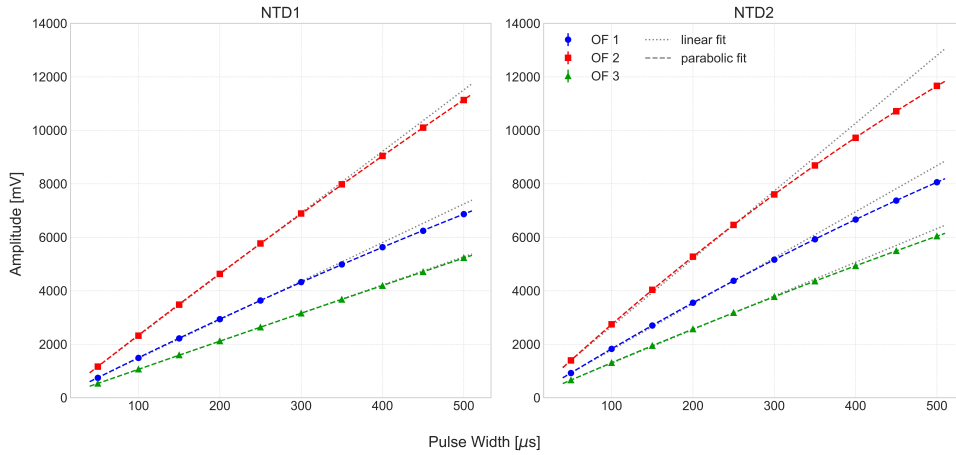}
  \caption{NTD amplitude vs. Pulse Width for both NTDs. The dashed lines are linear fits on the first six data points plus a physical constraint of the origin, such that the fit predicts no response in the NTD for no energy deposited in the crystal. 
  }
   \label{fig:pw_scan}
\end{figure*}

All data were processed using the optimum filtering technique \cite{Gatti1986}. 
The average noise power spectrum and mean pulse of the data acquired with the LED on the middle OF was used for both NTDs to build the transfer function applied to data acquired with the LED at each fiber. 
The filtered amplitude is further scaled to stabilize against the drifting baseline of the NTDs.
The LED pulse width (as a proxy for energy) is calibrated to operate within the linear range of the NTDs by performing a scan of the NTD amplitude over a range of LED pulse widths from 50 $\mu$s to 500 $\mu$s with a step of 50 $\mu$s. In Figure \ref{fig:pw_scan}, one can see the data deviated from the linear fit at 300 $\mu$s for all OFs. 
Thus for further commissioning, the LED pulse width was set to 200 $\mu$s to operate the NTDs safely in the linear regime. 
\begin{figure}[!htpb]
  \centering
 \begin{subfigure}{0.43\textwidth}
    \centering
    \includegraphics[width=\linewidth]{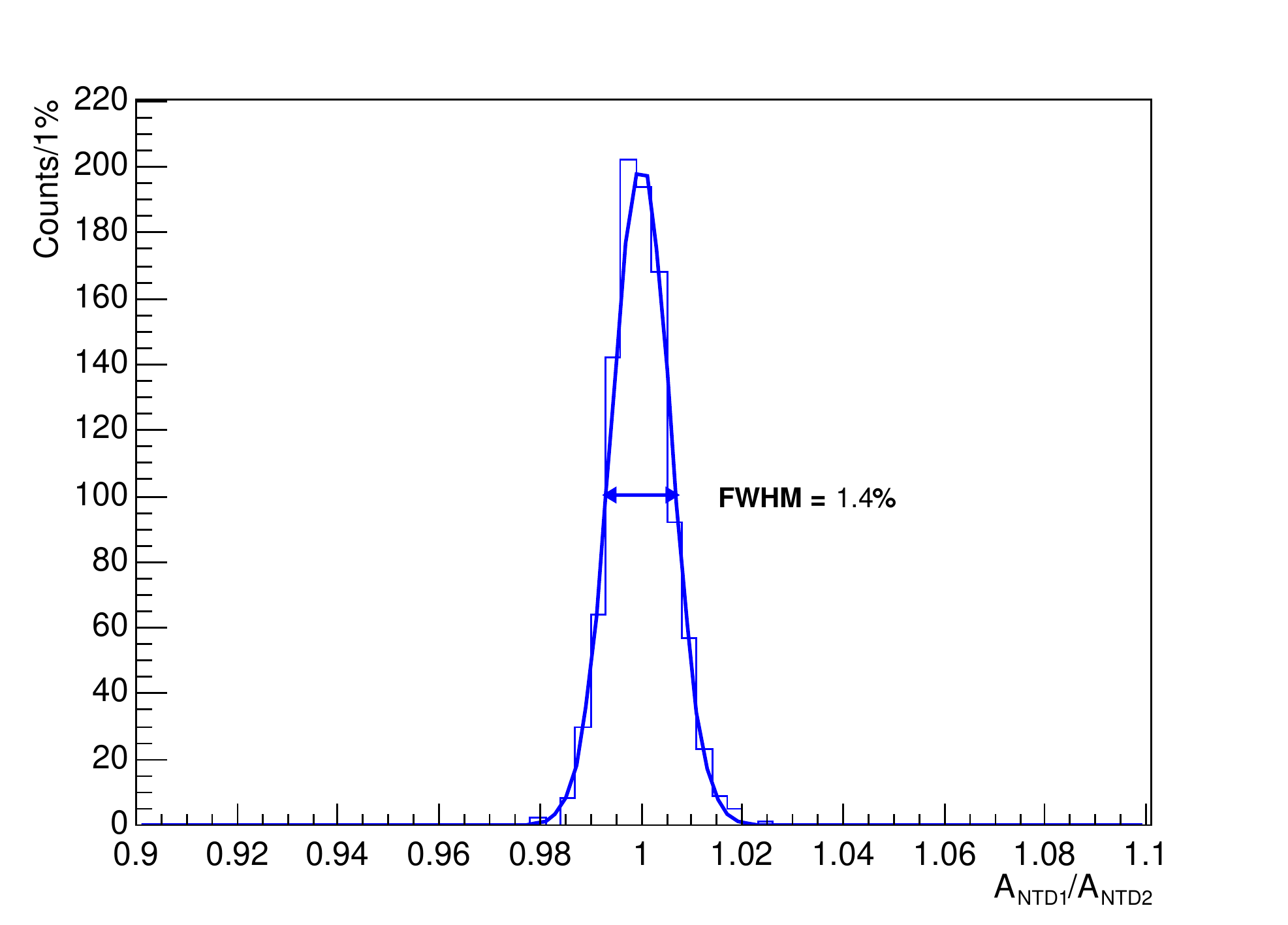}
    \caption{}
    \label{fig:amp_reso}
  \end{subfigure}
  \begin{subfigure}{0.43\textwidth}
    \includegraphics[width=\linewidth]{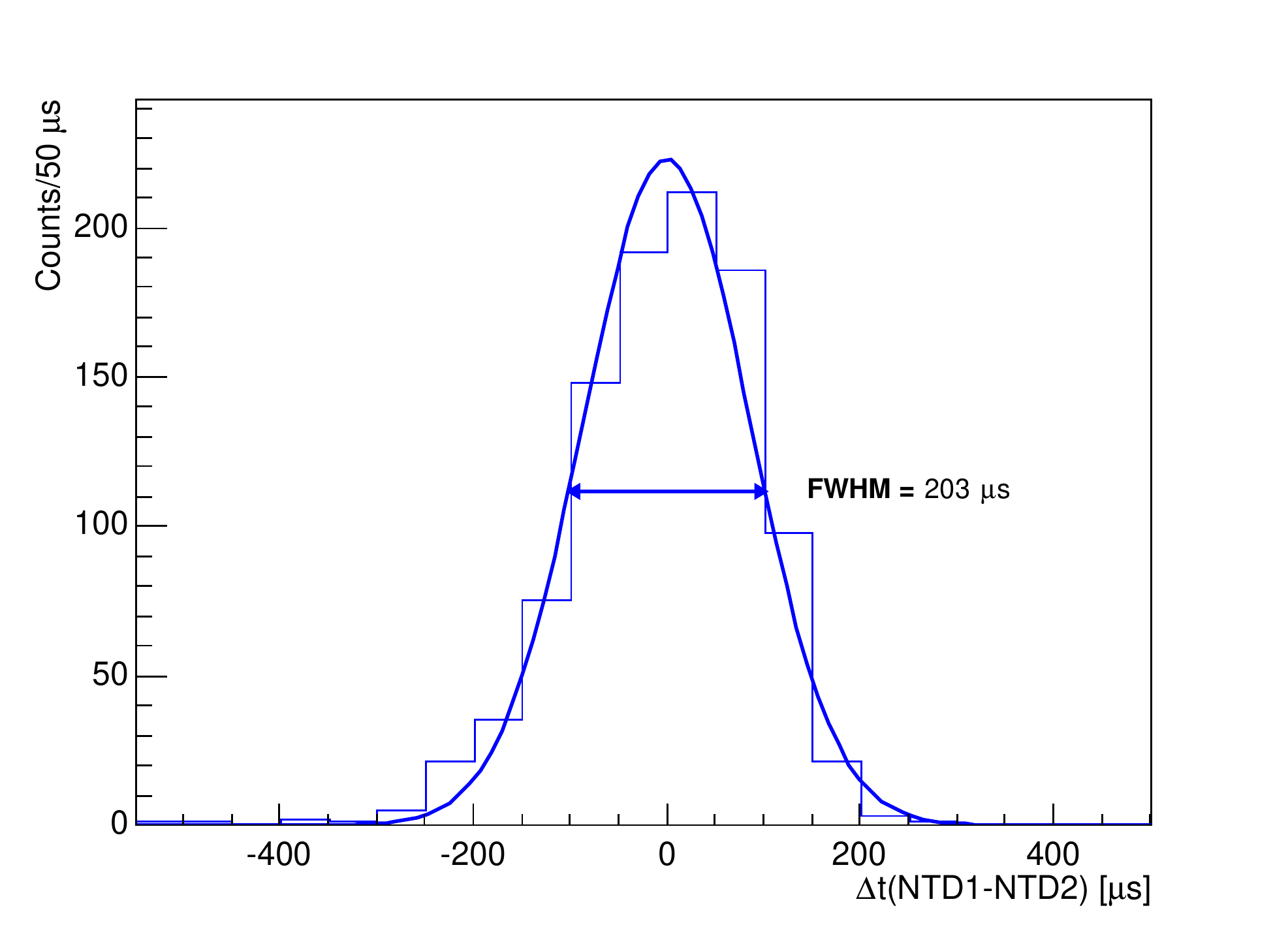}
    \caption{}
       \label{fig:timing}
  \end{subfigure}

  \caption{Fitted histograms displaying the timing (a) and energy sharing resolution (b). The double-sided arrow indicates the full-width-half-max of the Gaussian fit applied to the data. Pulses were collected with the LED on OF1.}

  \label{fig:resolution}
\end{figure}

\begin{figure*}[!t]
\centering
  \includegraphics[width=.7\textwidth]{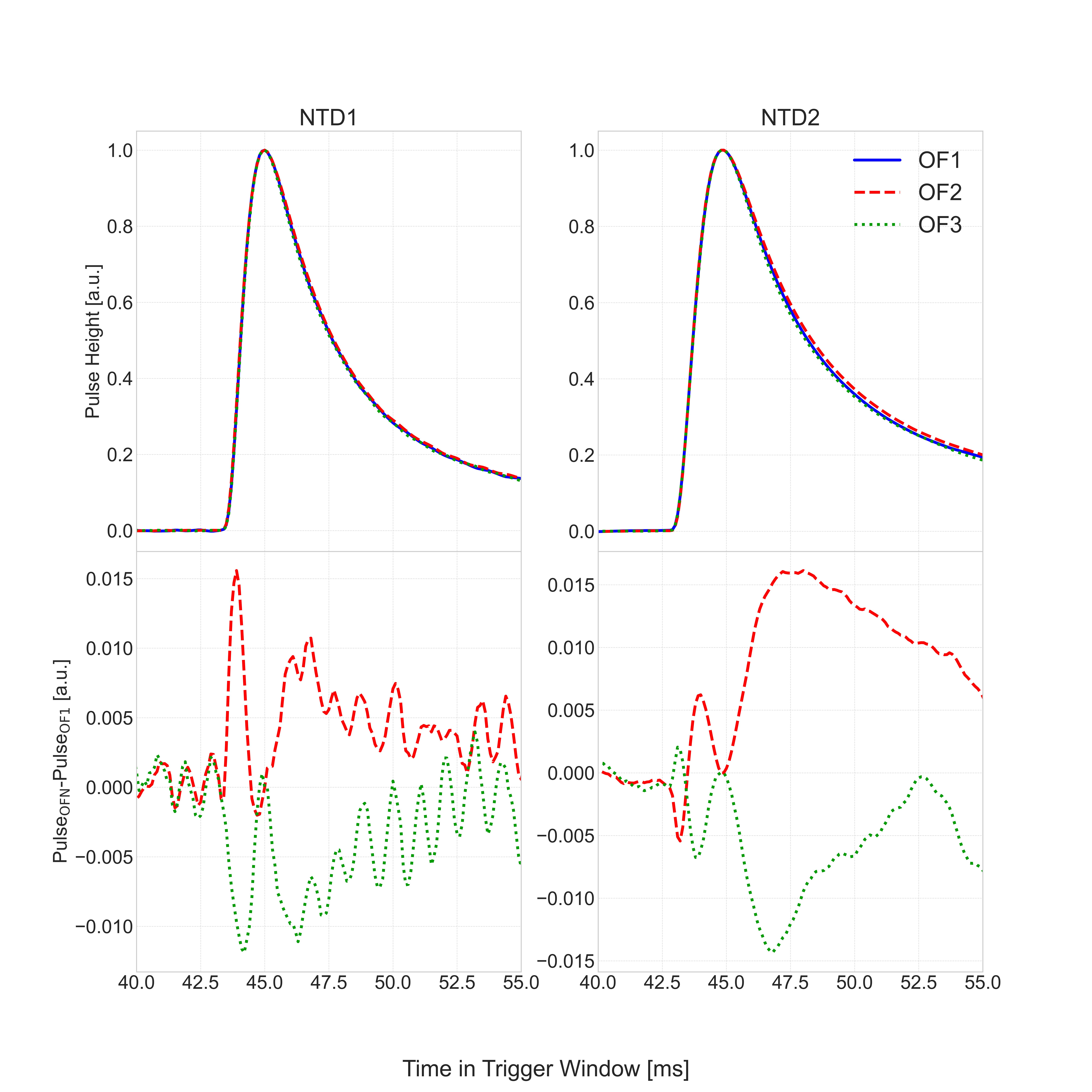}
  \caption{TOP: Amplitude normalized average pulses for both NTDs from each optical fiber (OF1-3). BOTTOM: Differences between normalized average pulses from each OF and OF1 to 
  highlight fiber-dependent pulse shape characteristics.
  }
   \label{fig:rise_time}
\end{figure*}

The amplitude-normalized average pulses we observed can be seen in the top plot of Figure \ref{fig:rise_time}, 
with differences in the pulse shape between the average for each OF accentuated by the plot on the bottom; 
however, further analysis is required to quantify the contributions due to position.
At this energy, the energy sharing resolution between the two NTDs was observed in \ref{fig:amp_reso} to be $\sigma(A_1/A_2)=0.59$\%, where the  the amplitude is a proxy for energy. This corresponds to a FWHM of 1.4\%. The mean of the distribution was scaled to 1 to calibrate differences in NTD response. 
Finally, we construct the timing resolution by defining the arrival time. This quantity is the time at which the pulse reaches 10\% of the maximum amplitude. Then, the timing resolution is defined as 1$\sigma$ for the distribution of the difference of the arrival times for NTD1 and NTD2 for pulses from OF1. 
As seen in Figure \ref{fig:timing}, we observed a timing resolution of $\sigma(t)=86\,\mu$s, which corresponds to a FWHM of 203 $\mu$s. As with the energy sharing resolution, the mean of the distribution was shifted to 0 to offset timing differences introduced
by the pulse shape differences and the readout.

\section{Conclusions and Future Work}
With commissioning complete, we will move forward with parameterizing the position-dependent effects on the pulse shape (arrival time, amplitude, pulse integral, etc.) for this 1D configuration. 
To this end, we are pursuing Monte Carlo simulations of ballistic phonon propagation in Geant4 \cite{Allison2006,AGOSTINELLI2003250,ALLISON2016186} and finite-element simulations to model thermal phonon propagation in COMSOL \cite{comsol64}.
We plan to also broaden types of radiation to include gamma, x-ray, and alpha sources in order to study single- and multi-site topologies in depth.

Once the analysis is complete for TeO$_2$ crystals, one can extend this methodology to other relevant bolometric crystals, such as Li$_2$MoO$_4$ used in CUPID. We are also preparing devices with TES's to perform fast sensor measurements to understand possible ballistic-to-thermal phonon conversion effects on the energy systematics.   

To systematically scan the crystal without repeated thermal cycles, integration of a cryo-compatible MEMS mirror optical system developed by N. Tabassum et. al \cite{tabassum2025broadbandopticalmodulationcontrol} is also planned. This steering mechanism will enable high-resolution mapping of the calorimetric response, dramatically improving spatial coverage and precision.
Finally, we plan the installation of an anti-vibrational suspension system to allow for stable data-taking conditions while the pulse tube cooler is active.

\newpage

\begin{acknowledgements}
We would like to thank Jesse Lopez for engineering and machine support, Vivek Singh and Bradford Welliver for their technical expertise and help with early conceptual designs, and the members of the Berkeley Weak Interactions Group for valuable discussions and stimulating intellectual atmosphere. 
This work was supported by the US Department of Energy (DOE) Office of Science, Office of Basic Energy Sciences under Contract No. DE-AC02-05CH11231, and by the DOE Office of Science, Office of High Energy and Nuclear Physics under Contract No. DE-FG02-08ER41551.
\end{acknowledgements}

\bibliographystyle{spphys}
\bibliography{ceres_concept}

@thesis{anisha_thesis,
  author       = {Yeddanapudi, Anisha },
  title        = {Understanding Energy Reconstruction in CUORE using Topological Analysis},
  note         = {Undergraduate honors thesis, {University California, Berkeley}},
  year         = {2024},
  month        = {May},
  url          = {}
}

@misc{comsol64,
  author = {{COMSOL AB}},
  title = {{COMSOL Multiphysics{\textregistered} v. 6.4}},
  year = {2025},
  url = {https://www.comsol.com},
  address = {Stockholm, Sweden}
}

@article{CUPID-Mo:2023vle,
    author = "Augier, C. and others",
    collaboration = "CUPID-Mo",
    title = "{The background model of the CUPID-Mo $0\nu \beta \beta $ experiment}",
    eprint = "2305.01402",
    archivePrefix = "arXiv",
    primaryClass = "hep-ex",
    doi = "10.1140/epjc/s10052-023-11830-2",
    journal = "Eur. Phys. J. C",
    volume = "83",
    number = "7",
    pages = "675",
    year = "2023"
}

@article{ALLISON2016186,
title = {Recent developments in Geant4},
journal = {Nuclear Instruments and Methods in Physics Research Section A: Accelerators, Spectrometers, Detectors and Associated Equipment},
volume = {835},
pages = {186-225},
year = {2016},
issn = {0168-9002},
doi = {https://doi.org/10.1016/j.nima.2016.06.125},
url = {https://www.sciencedirect.com/science/article/pii/S0168900216306957},
author = {J. Allison and others}
}

@article{Allison2006,
author = {Allison, J. and others},
year = {2006},
month = {02},
pages = {270-278},
title = {Geant4 Developments and Applications},
volume = {53},
journal = {IEEE Transactions on Nuclear Science},
doi = {10.1109/TNS.2006.869826}
}

@article{AGOSTINELLI2003250,
title = {Geant4—a simulation toolkit},
journal = {Nuclear Instruments and Methods in Physics Research Section A: Accelerators, Spectrometers, Detectors and Associated Equipment},
volume = {506},
number = {3},
pages = {250-303},
year = {2003},
issn = {0168-9002},
doi = {https://doi.org/10.1016/S0168-9002(03)01368-8},
url = {https://www.sciencedirect.com/science/article/pii/S0168900203013688},
author = {S. Agostinelli and others}
}

@Article{Kim2025,
author={Kim, W. T. and others},
title={A Study of Event Position Dependence for the AMoRE-II R\&D detectors with $\text{Li}_{2}\text{MoO}_{4}$ Crystal Absorbers},
journal={Journal of Low Temperature Physics},
year={2025},
month={Jan},
day={01},
volume={218},
number={1},
pages={83-91},
issn={1573-7357},
doi={10.1007/s10909-024-03246-3},
url={https://doi.org/10.1007/s10909-024-03246-3}
}

@article{CUORE:2021ctv,
    author = "Adams, D. Q. and others",
    collaboration = "CUORE",
    title = "{CUORE opens the door to tonne-scale cryogenics experiments}",
    eprint = "2108.07883",
    archivePrefix = "arXiv",
    primaryClass = "physics.ins-det",
    doi = "10.1016/j.ppnp.2021.103902",
    journal = "Prog. Part. Nucl. Phys.",
    volume = "122",
    pages = "103902",
    year = "2022"
}

@misc{tabassum2025broadbandopticalmodulationcontrol,
      title={Broadband Optical Modulation and Control at Millikelvin Temperatures}, 
      author={N. Tabassum and others},
      year={2025},
      eprint={2504.06995},
      archivePrefix={arXiv},
      primaryClass={physics.ins-det},
      url={https://arxiv.org/abs/2504.06995}, 
}

@article{Kadel_2016,
doi = {10.1088/1748-0221/11/03/T03004},
url = {https://doi.org/10.1088/1748-0221/11/03/T03004},
year = {2016},
month = {mar},
publisher = {},
volume = {11},
number = {03},
pages = {T03004},
author = {Kadel, R.W.},
title = {On the analytic estimation of radioactive contamination from degraded alphas},
journal = {Journal of Instrumentation}
}

@article{CUPID:2025avs,
    author = "Alfonso, K. and others",
    collaboration = "CUPID",
    title = "{CUPID, the Cuore upgrade with particle identification}",
    eprint = "2503.02894",
    archivePrefix = "arXiv",
    primaryClass = "physics.ins-det",
    doi = "10.1140/epjc/s10052-025-14352-1",
    journal = "Eur. Phys. J. C",
    volume = "85",
    number = "7",
    pages = "737",
    year = "2025",
    note = "[Erratum: Eur.Phys.J.C 85, 1346 (2025)]"
}

@Inbook{TES,
author="Irwin, K.D.
and Hilton, G.C.",
title="Transition-Edge Sensors",
bookTitle="Cryogenic Particle Detection",
year="2005",
publisher="Springer Berlin Heidelberg",
address="Berlin, Heidelberg",
pages="63--150",
isbn="978-3-540-31478-3",
doi="10.1007/10933596_3",
url="https://doi.org/10.1007/10933596_3"
}

@article{Haller1996,
author = {Haller, E. and others},
year = {1996},
month = {11},
pages = {115},
title = {Neutron Transmutation Depot (NTD) Germanium Thermistors for Submillimetre Bolometer Applications},
journal = {LBNL Report \#: LBNL-38912 },
volume = {388}
}

@Article{Gatti1986,
author={Gatti, E.
and Manfredi, P. F.},
title={Processing the signals from solid-state detectors in elementary-particle physics},
journal={La Rivista del Nuovo Cimento (1978-1999)},
year={1986},
month={Jan},
day={01},
volume={9},
number={1},
pages={1-146},
issn={1826-9850},
doi={10.1007/BF02822156},
url={https://doi.org/10.1007/BF02822156}
}

@article{CUPID:2022wpt,
    author = "Armatol, A. and others",
    collaboration = "CUPID",
    title = "{Toward CUPID-1T}",
    eprint = "2203.08386",
    archivePrefix = "arXiv",
    journal = "arXiv preprint arXiv:2203.08386",
    primaryClass = "nucl-ex",
    month = "mar",
    year = "2022"
}

@article{CUORE:2024ikf,
    author = "Adams, D. Q. and others",
    collaboration = "CUORE",
    title = "{Constraints on lepton number violation with the 2 tonne {\textperiodcentered} year CUORE dataset}",
    eprint = "2404.04453",
    archivePrefix = "arXiv",
    primaryClass = "nucl-ex",
    doi = "10.1126/science.adp6474",
    journal = "Science",
    volume = "390",
    number = "6777",
    pages = "1029--1032",
    year = "2025"
}

@misc{CUORE:topology,
    author = "Adams, D. Q. and others",
    collaboration = "CUORE",
  title = "{Topological Dependence of Energy Reconstruction in CUORE TeO2 Bolometers}",
  year = {2026},
note = "in preparation"
}

\end{document}